\documentclass[sigconf]{acmart}
\AtBeginDocument{%
  }

\setcopyright{acmlicensed}
\copyrightyear{2018}
\acmYear{2018}
\acmDOI{XXXXXXX.XXXXXXX}
\acmConference[Conference acronym 'XX]{Make sure to enter the correct
  conference title from your rights confirmation email}{June 03--05,
  2018}{Woodstock, NY}
\acmISBN{978-1-4503-XXXX-X/2018/06}

\usepackage{graphicx}
\usepackage{xcolor}
\usepackage{enumitem}
\usepackage{listings}

\definecolor{codebg}{RGB}{248,248,248}
\definecolor{keyword}{RGB}{0,0,255}
\definecolor{stringcolor}{RGB}{163,21,21}
\definecolor{keycolor}{RGB}{0,102,204}

\lstdefinelanguage{json}{
    basicstyle=\ttfamily\tiny,
    showstringspaces=false,
    breaklines=true,
    morestring=[b]",
    stringstyle=\color{stringcolor},
    literate=
     *{0}{{{\color{black}0}}}{1}
      {1}{{{\color{black}1}}}{1}
      {2}{{{\color{black}2}}}{1}
      {3}{{{\color{black}3}}}{1}
      {4}{{{\color{black}4}}}{1}
      {5}{{{\color{black}5}}}{1}
      {6}{{{\color{black}6}}}{1}
      {7}{{{\color{black}7}}}{1}
      {8}{{{\color{black}8}}}{1}
      {9}{{{\color{black}9}}}{1}
      {:}{{{\color{black}:}}}{1}
      {,}{{{\color{black},}}}{1}
      {\{}{{{\color{keycolor}\{}}}{1}
      {\}}{{{\color{keycolor}\}}}}{1}
}
\lstdefinestyle{jsonstyle}{
    backgroundcolor=\color{white},
    basicstyle=\selectfont\ttfamily\Normal,
    numbers=none,
    xleftmargin=0pt,
    xrightmargin=0pt,
    frame=none,
    breaklines=true,
    captionpos=b,
    language=json,
    showstringspaces=false
}

\copyrightyear{2025}
\acmYear{2025}
\setcopyright{cc}
\setcctype{by-nc-sa}
\acmConference[CIKM '25]{Proceedings of the 34th ACM International Conference on Information and Knowledge Management}{November 10--14, 2025}{Seoul, Republic of Korea}
\acmBooktitle{Proceedings of the 34th ACM International Conference on Information and Knowledge Management (CIKM '25), November 10--14, 2025, Seoul, Republic of Korea}\acmDOI{10.1145/3746252.3761629}
\acmISBN{979-8-4007-2040-6/2025/11}
\begin{document}

\title{YTCommentVerse: A Multi-Category Multi-Lingual YouTube Comment Corpus}

\author{Hridoy Sankar Dutta}
\affiliation{%
  \institution{Deakin University}
  \city{GIFT City}
  \country{India}}
\email{hridoy.dutta@deakin.edu.au}

\author{Biswadeep Khan}
\affiliation{%
  \institution{Stanford University}
  \city{California}
  \country{USA}}
\email{biswak@stanford.edu}

\renewcommand{\shortauthors}{Dutta and Khan}

\begin{abstract}
In this paper, we introduce YTCommentVerse, a large-scale multilingual and multi-category dataset of YouTube comments. It contains over 32 million comments from 178,000 videos contributed by more than 20 million unique users spanning 15 distinct YouTube content categories such as Music, News, Education and Entertainment. Each comment in the dataset includes video and comment IDs, user channel details, upvotes and category labels. With comments in over 50 languages, YTCommentVerse provides a rich resource for exploring sentiment, toxicity and engagement patterns across diverse cultural and topical contexts. This dataset helps fill a major gap in publicly available social media datasets particularly for analyzing video sharing platforms by combining multiple languages, detailed categories and other metadata.
\end{abstract}



\maketitle

\section{Introduction}
YouTube has emerged as one of the most influential digital platforms, hosting billions of videos and engaging users across the globe in a wide range of interactions, from entertainment and education to politics and activism \cite{rieder2023making}. With over 2 billion monthly active users \cite{shepherd2025youtube} and content that spans all major languages and cultures \cite{nbc2023youtube}, YouTube comments offer a unique lens into how people respond to and interact with video content. 

A substantial number of studies have investigated social media platforms for a range of problems related to social network structures, spam and coordinated behaviours mainly on Twitter \cite{alsaleh2014tsd,gupta2014tweetcred,giatsoglou2015retweeting,shah2014spotting,chavoshi2016debot}, Facebook \cite{aljabri2023machine, perrotta2021behaviours,ernala2020well} and Reddit \cite{sakketou2022factoid,nakamura2019r,corsi2024crowdsourcing}. Twitter and Reddit, in particular, have been the dominant source of data for studying online manipulation due to the relatively open API that has enabled researchers to extract and construct large-scale datasets from these platforms. In contrast, YouTube being the second-most visited website globally has received comparatively less attention in the academic research community. This is due in part to the challenges associated with collecting and curating YouTube data. Consequently, most existing YouTube comment datasets are either narrow in scope, limited in size or constrained by the availability of only a few metadata fields.

\begin{figure}[!b]
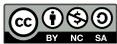

\centering
\begin{minipage}[t]{0.5\columnwidth}
\raggedright
\begin{lstlisting}[style=jsonstyle]
{
    "videoID": "ab9fe...",
    "commentID": "488b2...",
    "commentorName": "b65...",
    "commentorChannelID": "2f1 ...",
    "comment": "ich fand den 
    Handelwecker am besten",
    "votes": 2,
    "originalChannelID": "2f13...",
    "category": "entertainment"
}
\end{lstlisting}
\end{minipage}%
\hspace{0.02\columnwidth}%
\begin{minipage}[t]{0.45\columnwidth}
\null\centering
\begin{tabular}{@{}lc@{}}
\toprule
Metric & Count \\ \midrule
\# videos & 178,027 \\
\# comments & 32,236,173 \\
\# commentors & 20,568,637 \\
\# categories & 15 \\
\bottomrule
\end{tabular}
\end{minipage}
\caption{A sample YouTube comment and data statistics.}
\label{fig:youtube_comments_statistics}
\end{figure}

To address this gap, we present YTCommentVerse, a large-scale, multilingual and multi-category dataset of over 32 million YouTube comments drawn from 178,000 videos across 15 distinct content categories (cf. Figure \ref{fig:youtube_comments_statistics}). The dataset includes rich metadata from comment IDs, user channel identifiers, upvotes and video category labels and spans more than 50 languages, making it one of the most diverse and comprehensive YouTube comment corpora to date. It is important to note that all Personally Identifiable Information (PII) has been redacted in the released dataset. In Table \ref{tab:yt_datasets_detailed}, we show a comparison of existing YouTube datasets with YTCommentVerse. While prior datasets offer limited metadata, YTCommentVerse includes over 30M comments and rich metadata making it the most comprehensive resource for YouTube comment analysis.

\begin{table}[!t]
\centering
\small
\caption{\small Comparison of YouTube comment datasets with available metadata fields.}
\begin{tabular}{|l|c|p{4cm}|}
\hline
\textbf{Dataset} & \textbf{\#} & \textbf{Key Fields} \\
\hline
YouTube18K \cite{atifaliak_youtube_2022} & $\sim$18K & comment, sentiment \\
COVID-YT \cite{seungguini_youtube_2021} & $\sim$24K & url, title, channel, commentText \\
YTSpam \cite{waheed2023youtube} & $\sim$1.9K & commentID, commentorName, commentDate, comment, videoName \\
YouTube-180K \cite{breadlicker45_youtube_comments_180k} & $\sim$180K & comment \\
\textbf{YTCommentVerse} & \textbf{30M+} & \textbf{comment, videoID, commentID, commentorName, commentorChannelID, upvotes, originalChannelID, category} \\
\hline
\end{tabular}
\label{tab:yt_datasets_detailed}
\end{table}

\begin{table*}[!htbp]
\caption{Language Distribution of Comments}
\label{tab:language_distribution}
\setlength{\tabcolsep}{3pt}
\begin{tabular}{@{}lllllllllllllllllllll@{}} 
\toprule
\textbf{ISO} & \textbf{\#} & \textbf{\%} &
\textbf{ISO} & \textbf{\#} & \textbf{\%} &
\textbf{ISO} & \textbf{\#} & \textbf{\%} &
\textbf{ISO} & \textbf{\#} & \textbf{\%} &
\textbf{ISO} & \textbf{\#} & \textbf{\%} &
\textbf{ISO} & \textbf{\#} & \textbf{\%} &
\textbf{ISO} & \textbf{\#} & \textbf{\%} \\
\midrule
en    & 8.9M  & 22.0 & es    & 4.4M  & 10.8 & ko    & 3.8M  & 9.4  & pt    & 3.7M  & 9.0  & und   & 2.3M  & 5.6  & th    & 2.3M  & 5.6  & ar    & 1.2M  & 3.1  \\
id    & 1.2M  & 2.9  & so    & 1.1M  & 2.8  & de    & 1.1M  & 2.7  & it    & 750K  & 1.8  & fr    & 707K  & 1.7  & tl    & 648K  & 1.6  & sw    & 637K  & 1.6  \\
bn    & 594K  & 1.5  & vi    & 550K  & 1.4  & et    & 537K  & 1.3  & nl    & 500K  & 1.2  & tr    & 480K  & 1.2  & sl    & 435K  & 1.1  & ro    & 424K  & 1.0  \\
fi    & 415K  & 1.0  & ca    & 413K  & 1.0  & af    & 390K  & 1.0  & cy    & 341K  & 0.8  & no    & 303K  & 0.7  & hr    & 253K  & 0.6  & ru    & 241K  & 0.6  \\
pl    & 241K  & 0.6  & da    & 209K  & 0.5  & fa    & 202K  & 0.5  & sv    & 198K  & 0.5  & hu    & 193K  & 0.5  & sq    & 189K  & 0.5  & sk    & 152K  & 0.4  \\
hi    & 110K  & 0.3  & lt    & 108K  & 0.3  & ja    & 105K  & 0.3  & cs    & 103K  & 0.3  & lv    & 52.1K & 0.1  & ur    & 50.6K & 0.1  & bg    & 24.8K & 0.1  \\
zh-cn & 21.3K & 0.1  & te    & 16.3K & 0.04 & mr    & 16.2K & 0.04 & uk    & 16.0K & 0.04 & mk    & 15.6K & 0.04 & ne    & 9.3K  & 0.02 & el    & 7.6K  & 0.02 \\
ml    & 5.2K  & 0.01 & zh-tw & 4.9K  & 0.01 & ta    & 4.0K  & 0.01 & gu    & 3.4K  & 0.01 & pa    & 2.3K  & 0.01 & he    & 1.8K  & 0.005 & kn    & 1.2K  & 0.003 \\
\bottomrule
\end{tabular}
\end{table*}

\section{Building YTCommentVerse}
\subsection{Data collection}
 We use the SocialBlade\footnote{\url{https://socialblade.com/}} ranking to get the top YouTube channels based on the most number of subscribers for each categories. The channels in SocialBlade are categorized into 15 different categories: `Autos \& Vehicles' \textit{(autos)}, `Comedy', `Education', `Entertainment', `Film', `Gaming', `Science \& Technology' \textit{(tech)}, `Howto \& Style' \textit{(howto)}, `Music', `News \& Politics' \textit{(news)}, `Nonprofit \& Activism' \textit{(nonprofit)}, `People \& Blogs' \textit{(people)}, `Pets \& Animals' \textit{(animals)}, `Sports' and `Travel'.

We developed custom web scrapers to collect video information from each YouTube channel. First, we create a list of videos posted by a YouTube channel. Second, we collect the video metadata and the comments along with its metadata that have been posted on a video. Note that we anonymize all Personally Identifiable Information (PII) in our dataset.
\subsection{Data characteristics}
Each entry in the dataset is related to one comment for a specific YouTube video in the related category with the following columns: videoID, commentID, commentorName, 
 commentorChannelID, comment, upvotes, originalChannelID, category. Each field is explained below:
 \begin{enumerate}[leftmargin=*]
     \item \textbf{videoID:} represents the video ID in YouTube.
     \item \textbf{commentID:} represents the comment ID.
     \item \textbf{commentorName:} represents the name of the commentor.
     \item \textbf{commentorChannelID:} represents the ID of the commentor.
     \item \textbf{comment:} represents the comment text.
     \item \textbf{Upvotes:} represents the upvotes received by that commment.
     \item \textbf{originalChannelID:} represents the original channel ID who posted the video.
     \item \textbf{category:} represents the category of the YouTube video.
 \end{enumerate}

\section{Data analysis}
\begin{figure}[!htbp]
    \centering
    \includegraphics[width=\linewidth]{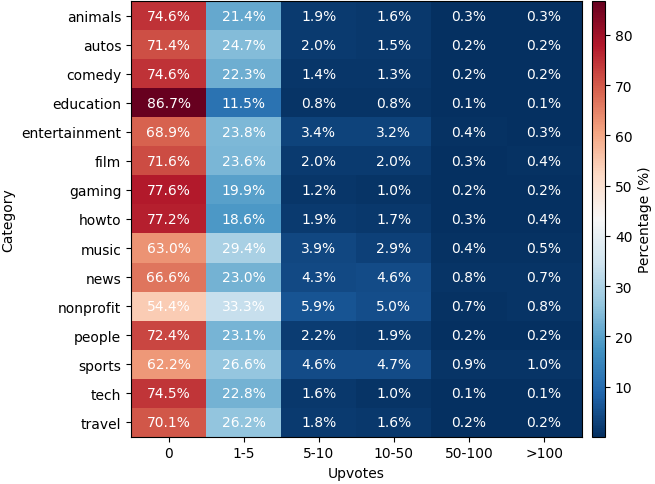}
    \caption{Upvotes distribution for YouTube categories.}
    \label{fig:upvotes_dist}
    \end{figure}

We analyze the YTCommentVerse dataset across multiple dimensions to understand user engagement patterns and communication behaviors on YouTube: \textit{languages}, \textit{upvotes} and \textit{comment length}.
\begin{itemize}[leftmargin=*]
    \item \textbf{Language distribution:} Table \ref{tab:language_distribution} presents the distribution of comments across different languages from the YTCommentVerse dataset, identified by their ISO 639-1 codes. English dominates the dataset with 22\% of the comments, followed by Spanish (10.8\%), Korean (9.4\%) and Portuguese (9.0\%). We observe the presence of high-resource and low-resource languages that showcase the linguistic diversity in the dataset. The support for low-resource languages will significantly advance the development of large language models (LLMs).
   
    \item \textbf{Upvotes distribution:} Upvotes reflects the crowdsourced evaluation of audience engagement and is helpful to identify what content a viewer finds insightful or worth discussing. Figure \ref{fig:upvotes_dist} shows the upvotes distribution for YouTube categories. We observe that most comments across all categories receive little to no engagement, often fewer than five upvotes. 
    
    \item \textbf{Comment Length:} We perform the comment length analysis across 15 categories to user engagement depth and communication patterns. The Nonprofit category generates the most detailed discussions, with an average comment length of 111.7 characters. Comedy exhibits the highest variation in comment length ($\sigma = 1604.7$), suggesting different types of engagement within this category. 
\end{itemize}
\section{Benchmark Results: Sentiment analysis, Text embeddings and Toxicity analysis}
\begin{figure*}[!ht]
    \centering
    \includegraphics[width=0.95\linewidth]{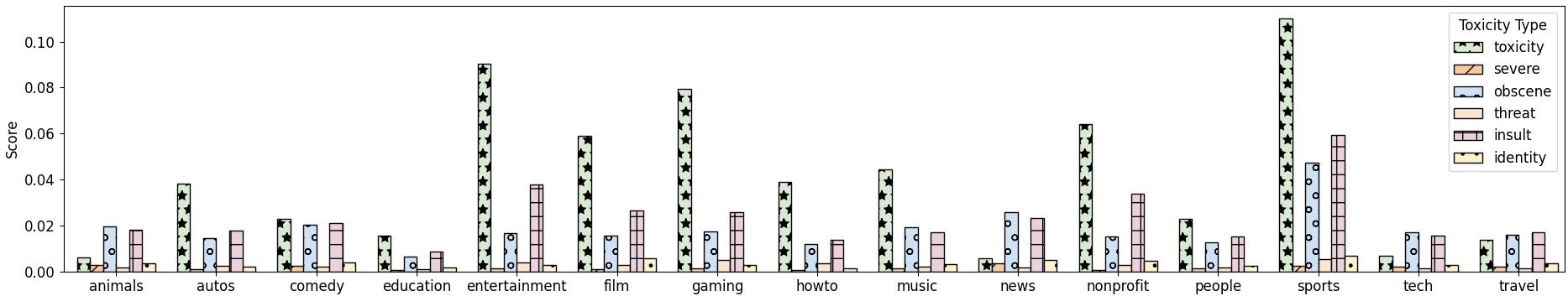}
    \caption{Toxicity Score for YouTube categories.}
    \label{fig:toxicity}
\end{figure*}

To assess the nature of user engagement across various YouTube categories, we performed multiple benchmark experiments: \textit{sentiment analysis}, \textit{text embeddings} and \textit{toxicity analysis} on the YouTube comment dataset. 
\begin{itemize}[leftmargin=*]
    \item \textbf{Sentiment analysis:} The sentiment analysis across 15 categories reveals distinct patterns in user engagement and emotional response. Music emerges as the most positively received category, with a mean sentiment of 0.142. Music also shows the highest sentiment variability ($\sigma = 0.299$), indicating polarized user opinions. Overall, we observe that the sentiment distribution suggests positive user engagement across the platform’s content categories. 
    
    We also show the relationship between sentiment polarity and upvotes across various content categories on YouTube in Figure \ref{fig:upvotes}. Most comments have neutral sentiment suggesting that a large portion of user engagement is associated with emotionally neutral content. Both positive and negative sentiments receive high upvotes in some cases, indicating that sentiment alone is not a strong predictor of popularity. 
    \begin{figure}[!htbp]
    \centering
    \includegraphics[width=\linewidth]{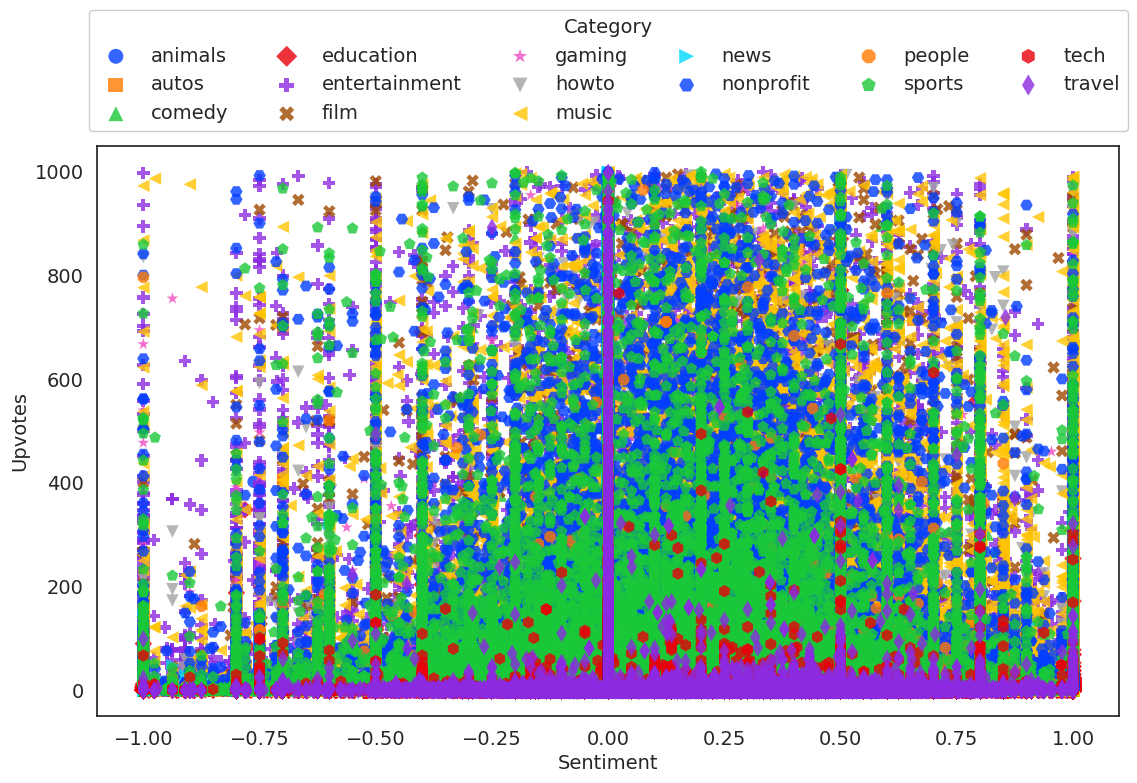}
    \caption{Sentiment vs. Upvotes distribution for YouTube categories.}
    \label{fig:upvotes}
    \end{figure}

    \item \textbf{Text embeddings:} Figure \ref{fig:embedding_plot}  shows the embedding of YouTube video comments, where each point represents a comment color-coded according to the video category (e.g., music, gaming, education, etc.). We performed a dimensionality reduction analysis on YouTube video comment embeddings to visualize their distribution across content categories. We sampled 100000 random comments from YTCommentVerse uniformly to ensure fair representation across the dataset. We applied UMAP and applied DBSCAN clustering to isolate the main cluster of dense activity by removing the outliers. 
    
    We observed that categories such as music, gaming and education formed relatively tight and coherent clusters, indicating that user comments in these areas share strong semantic similarities. Interestingly, categories like comedy, entertainment and tech showed more  overlap, suggesting broader topical ranges in audience interactions. 
    \begin{figure}
    \centering
    \includegraphics[width=\linewidth]{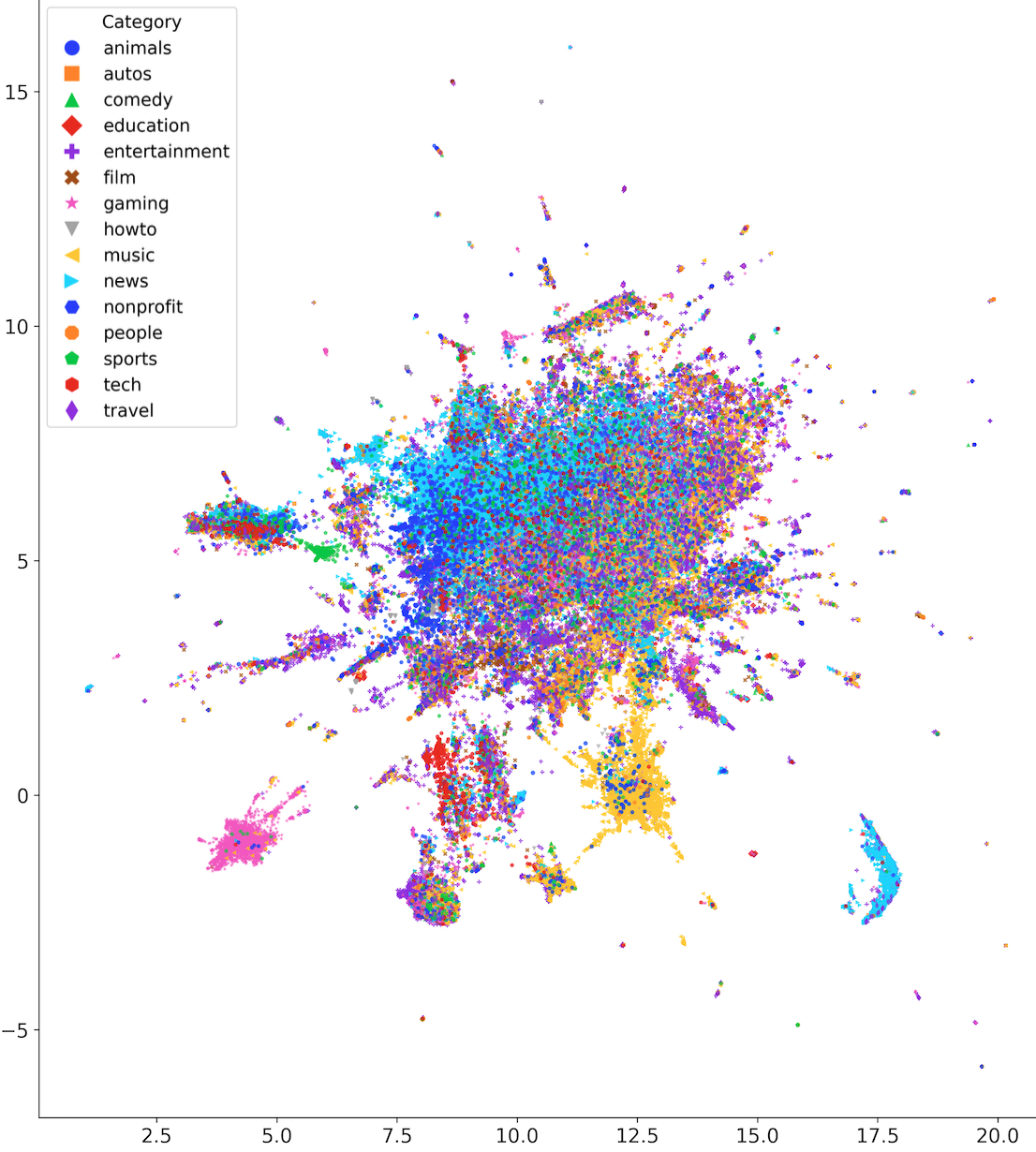}
    \caption{A high-dimensional embedding of YouTube video comments using UMAP.}
    \label{fig:embedding_plot}
    \end{figure}
    \item \textbf{Toxicity analysis:} Figure~\ref{fig:toxicity} illustrates the distribution of toxicity in YouTube comments across various video categories using six dimensions from the Perspective API\footnote{\url{https://perspectiveapi.com/}}: \textit{toxicity}, \textit{severe toxicity}, \textit{obscene}, \textit{threat}, \textit{insult} and \textit{identity}. Comments on entertainment, gaming and sports videos show the highest levels of toxicity, particularly in terms of general and severe toxicity, insults and obscene language, possibly due to competitive communities. In contrast, categories such as education, animals and travel exhibit much lower toxicity, reflecting more constructive user engagement. Identity-based toxicity remains low overall but is slightly more present in entertainment and gaming suggesting possible targeted remarks. 
\end{itemize} 

\section{Research Opportunities Using YTCommentVerse}

YTCommentVerse opens up a wide range of possibilities for research and development across natural language processing, computational social science and content moderation. Below, we highlight several use cases that demonstrate where YTCommentVerse can have a significant impact.
\begin{itemize}[leftmargin=*]
  \item \textbf{Cross-Cultural Discourse Analysis:} The dataset enables exploration of how global users engage with content differently depending on region, language and topical domain. This facilitates studies in sociolinguistics, cultural studies and digital anthropology.
  \item \textbf{Content Moderation:} The combination of comment text and engagement metrics such as upvotes offers a valuable resource for developing and validating models to detect toxic, obscene or other harmful content in online discussions. In addition to this, moderators can use it by identifying posts with unusual behaviors in upvotes that may signal harmful or manipulative content.
  \item \textbf{Fine-Tuning Large Language Models:} Given the scale of YTCommentVerse, the dataset can help adapt large language models to better understand different engagement patterns present in comments of video-sharing platforms leading to better tools for content creators,  and platform moderators.
  \item \textbf{Bias and Fairness Audits:} YTCommentVerse supports evaluation of algorithmic fairness by allowing researchers to examine differential treatment of comments across languages, regions or user types. This is particularly relevant for assessing harms while deploying large language models.
\end{itemize}

\section{Guidelines related to YTCommentVerse}
\subsection{FAIR-Guiding Principles}
To ensure the YTCommentVerse dataset adheres to the FAIR data principles, several measures have been taken. For Findability and Accessibility, a smaller version of the dataset has been publicly hosted on the Hugging Face Hub\footnote{\url{10.57967/hf/3602}}. Due to the scale of the dataset, YTCommentVerse (previously known as YT-30M) dataset could only be obtained by directly emailing the authors of this paper. This makes it findable through the Hub's search functionality and directly accessible using its unique identifier. To make the YTCommentVerse dataset Interoperable and Re-usable, it is structured for seamless integration with the Hugging Face datasets library. The dataset card on Hugging Face serves as comprehensive documentation similar to a README file, detailing the data structure, fields, potential uses and licensing information, which is crucial for optimizing its re-use by the community.
\subsection{Ethical Considerations}
We only collected data from publicly accessible YouTube videos and comment sections, which is permitted under applicable laws\footnote{\url{https://techcrunch.com/2022/04/18/web-scraping-legal-court/}}. We did not seek individual consent  each commenter as contacting millions of users would be impractical. Our assumption while collecting the dataset is that users are aware that comments posted publicly on YouTube are visible to anyone. All personally identifiable information (PII) has been anonymized on YTCommenVerse.

\section{Conclusion, Limitations and Release}
YouTube as a social media platform that has been largely overlooked in academic research. To  the  best  of  our  knowledge, YTCommentVerse is the first large-scale multilingual and multi-category dataset built from over 30 million YouTube comments sourced from 178,000 videos and contributed by more than 20 million unique users spanning 15 content categories. 

While YTCommentVerse provides a large-scale and diverse corpus for analyzing user discourse across YouTube video categories, several limitations must be considered. First, the dataset includes only publicly available comments, which introduces a bias toward highly active or engaged subscribers and may not reflect broader audience sentiment. Second, the dataset is non-temporal limiting its use for various temporal modeling tasks. Third, the dataset is inherently text-centric and does not include video content or thumbnails limiting the contextual interpretation of the comments. Upvotes provides only a partial view of user engagement, as it excludes dislikes, replies or other types of interactions.

A smaller version of the YTCommentVerse dataset called YT-100K is already available on the Hugging Face platform\footnote{\url{https://huggingface.co/datasets/hridaydutta123/YT-100K}}. At the time of writing this paper, the dataset already had more than 300 downloads. YTCommentVerse is currently uploaded to Zenodo\footnote{\url{https://zenodo.org/records/15678816}} in the form of an SQLite database with the columns mentioned in Section 2.2. The authors encourage researchers working in the domain of Natural Language Processing and Social Network Analysis to perform various interesting analyses and modeling on this dataset.
\section{GenAI Usage Disclosure}
No generative AI tools were used in the preparation of this manuscript.
\bibliographystyle{ACM-Reference-Format}
\bibliography{sample-base}
\end{document}